\documentclass{emulateapj} 
\usepackage{apjfonts}

\usepackage{amssymb}
\usepackage{amsmath}
\newcommand{\be}{\begin{equation}}
\newcommand{\ee}{\end{equation}}

\shorttitle{Formation of Old Globular Clusters in LCDM} 
\shortauthors{Trenti et al.}

\begin{document}


\title{The relative and absolute ages of old globular clusters in the $\Lambda$CDM framework}

\author{Michele Trenti\altaffilmark{1},Paolo Padoan\altaffilmark{2}, Raul Jimenez\altaffilmark{2,3}}

\altaffiltext{1}{School of Physics, The University of Melbourne, VIC
  3010, Australia}
\altaffiltext{2}{ICREA \& ICC, University of Barcelona, Marti i Franques 1, 08028 Barcelona, Spain}
\altaffiltext{3}{Institute for Applied Computational Science, Harvard University, MA 02138, USA}
\email{mtrenti@unimelb.edu.au} 

%

\begin{abstract}

  Old Globular Clusters (GCs) in the Milky Way have ages of about $13$
  Gyr, placing their formation time in the reionization epoch. We
  propose a novel scenario for the formation of these systems based on
  the merger of two or more atomic cooling halos at high-redshift
  ($z>6$). First generation stars are formed as an intense burst in
  the center of a minihalo that grows above the threshold for hydrogen
  cooling (halo mass $M_h\sim10^8~\mathrm{M_{\sun}}$) by undergoing a
  major merger within its cooling timescale ($\sim 150$
  Myr). Subsequent minor mergers and sustained gas infall bring new
  supply of pristine gas at the halo center, creating conditions that
  can trigger new episodes of star formation. The dark-matter halo
  around the GC is then stripped during assembly of the host galaxy
  halo. Minihalo merging is efficient only in a short redshift window,
  set by the $\Lambda CDM$ parameters, allowing us to make a strong
  prediction on the age distribution for old GCs. From cosmological
  simulations we derive an average merging redshift
  $\langle z\rangle=9$ and narrow distribution $\Delta z=2$, implying
  average GC age $\langle t_{age}\rangle=13.0\pm0.2~\mathrm{Gyr}$
  including $\sim0.2$ Gyr of star formation delay.  Qualitatively, our
  scenario reproduces other general old GC properties (characteristic
  masses and number of objects, metallicity versus galactocentric
  radius anticorrelation, radial distribution), but unlike age, these
  generally depend on details of baryonic physics. In addition to
  improved age measurements, direct validation of the model at
  $z\sim10$ may be within reach of ultradeep gravitationally lensed
  observations with the \emph{James Webb Space Telescope}.

\end{abstract}


\section{Introduction}\label{sec:intro}Globular Clusters (GCs) are compact stellar systems with
characteristic mass $\sim10^5~\mathrm{M_{\odot}}$ and radius of a few
pc \citep{heggie_hut_book}. Their ubiquitous
presence around galaxies and their old stellar populations
make them tools to investigate early star formation, assembly
history of host galaxies, and cosmological models
\citep{katz2013,brodie2014}. For example, GC ages were used to
constrain the age of the Universe \citep{Jimenez96,krauss2003_age},
providing early independent support to the concordance
$\Lambda CDM$ model \citep{planck}.

Yet, GC formation remains a debated topic with the lack of an
established scenario matching all observations. Early
proposals identified the high-$z$ Jeans mass 
($\approx10^6~\mathrm{M_{\odot}}$) with a protoglobular cloud
(\citealt{peebles1968}). More recently, high-$z$ GC formation as been
proposed within dark-matter (DM) halos of mass
$M_h\lesssim10^8~\mathrm{M_{\odot}}$, with cooling driven by $H_2$
(\citealt{padoan1997}), as result of shocks induced by a reionization
front \citep{cen01}, or as purely baryonic systems because stream
velocity displaced gas from its parent halo \citep{naoz2014}. Other
scenarios focus at somewhat lower redshift, ranging from
cooling-induced fragmentation of (proto)-galaxies \citep{fall1985}, to
formation during galaxy merging/interactions
(\citealt{ashman1992,muratov2010,li2014}) or within high-density
regions of galactic disks
\citep{kravtsov2005,kruijssen2014_review}. Analogies with
today's young massive clusters have also been proposed (e.g.,
\citealt{bastian2013}).

The lack of consensus on GC formation may indicate 
multiple formation mechanisms. In fact, the
oldest GCs have essentially uniform ages centered at $\sim12.8$ Gyr
and spread $\sim5\%$ (comparable to relative age uncertainty), but
wide range of metallicities: from 1\% to $1/3$ the solar value with
high metallicity systems preferentially at small galactocentric
radii. In contrast, younger systems show a well defined
age-metallicity anti-correlation and higher metallicity at larger
galactocentric radii
\citep{forbes2010_age_metal,marin-franch2009_ages}.

In this \emph{Letter} we aim at predicting the probability
distribution function of ages for the sub-population of ``old'' GCs
starting from DM halo assembly and merging, which are fully determined
(in the statistical sense) by fixing the $\Lambda CDM$ parameters. Our
approach parallels studies of evolution of galaxy properties
with redshift, such as luminosity/stellar mass functions
and clustering, all linked to halo assembly
\citep{trenti2010_lf,lacey2011,tacchella2013,behroozi2013}.

We propose that old GCs formed via major merging of DM minihalos
($M_{DM}\sim10^8~\mathrm{M_{\sun}}$) that are gas rich (no previous
star formation) and metal-enriched through outflows originating from
nearby (proto)galaxies. First generation stars form during the
merger-triggered burst, while subsequent gas infall through minor
mergers can lead to multiple populations. Since the halo merger rate
is approximately constant per unit redshift (both for major and minor
events; \citealt{fakhouri2010}), there is only a short window of
opportunity for this mechanism. The redshift needs to be low enough
($z\lesssim 15$) so that minihalos are relatively common, but not too
low ($z\gtrsim 6$), otherwise the merger rate drops.

\begin{figure}
\begin{center}
\includegraphics[scale=0.34]{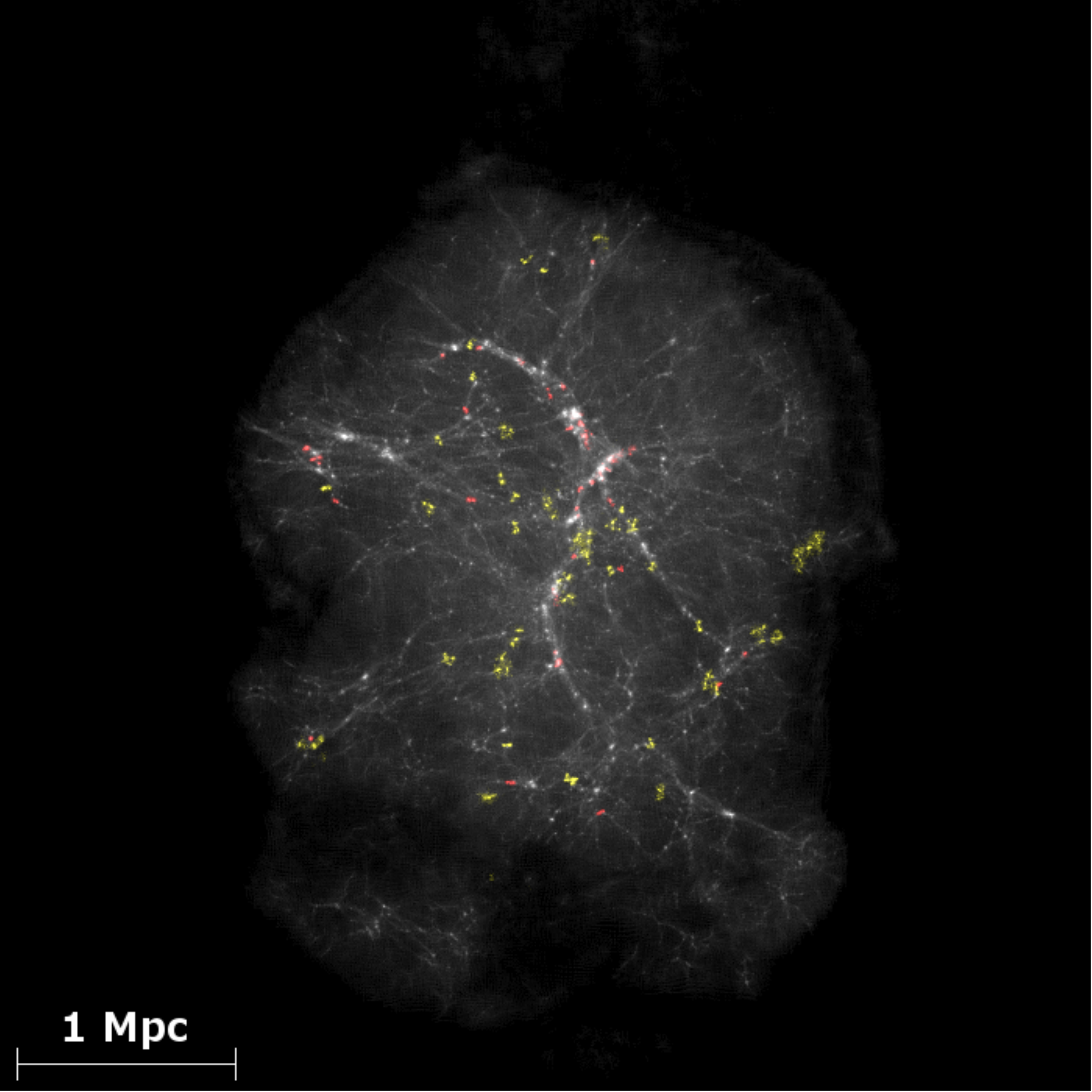}
\end{center}
\caption{Snapshot at $z=10$ ($t_{lookback}\sim13.3$ Gyr) of the region
  that by $z=0$ will collapse to the halo of mass
  $8\times 10^{11}~\mathrm{M_{\odot}}$ shown in the right panels of
  Figure~\ref{fig:rendering}. A linear overdensity containing hundreds
  of collapsing/collapsed minihalos is present. Regions involved in GC
  formation through minihalo mergers are highlighed in red (past major
  merger) and yellow (future merger). A movie capturing redshift evolution 
  is available in the online edition.}\label{fig:MWinfancy}
\end{figure}

\section{GC formation through minihalo-minihalo mergers}\label{sec:model}We propose that old GCs form when {{a host}} galaxy like the MW is in the
earliest stages of its assembly and lacks a well defined disk
structure. Its Lagrangian region is still a moderate linear
overdensity at $z\sim 10$ containing hundreds of minihalos that are
being hierarchically assembled, form stars, pollute their neighbours
with metals, and merge to build up a massive system by $z=0$
(Figure~\ref{fig:MWinfancy}).

\begin{figure*}
\begin{center} 
\includegraphics[scale=0.39]{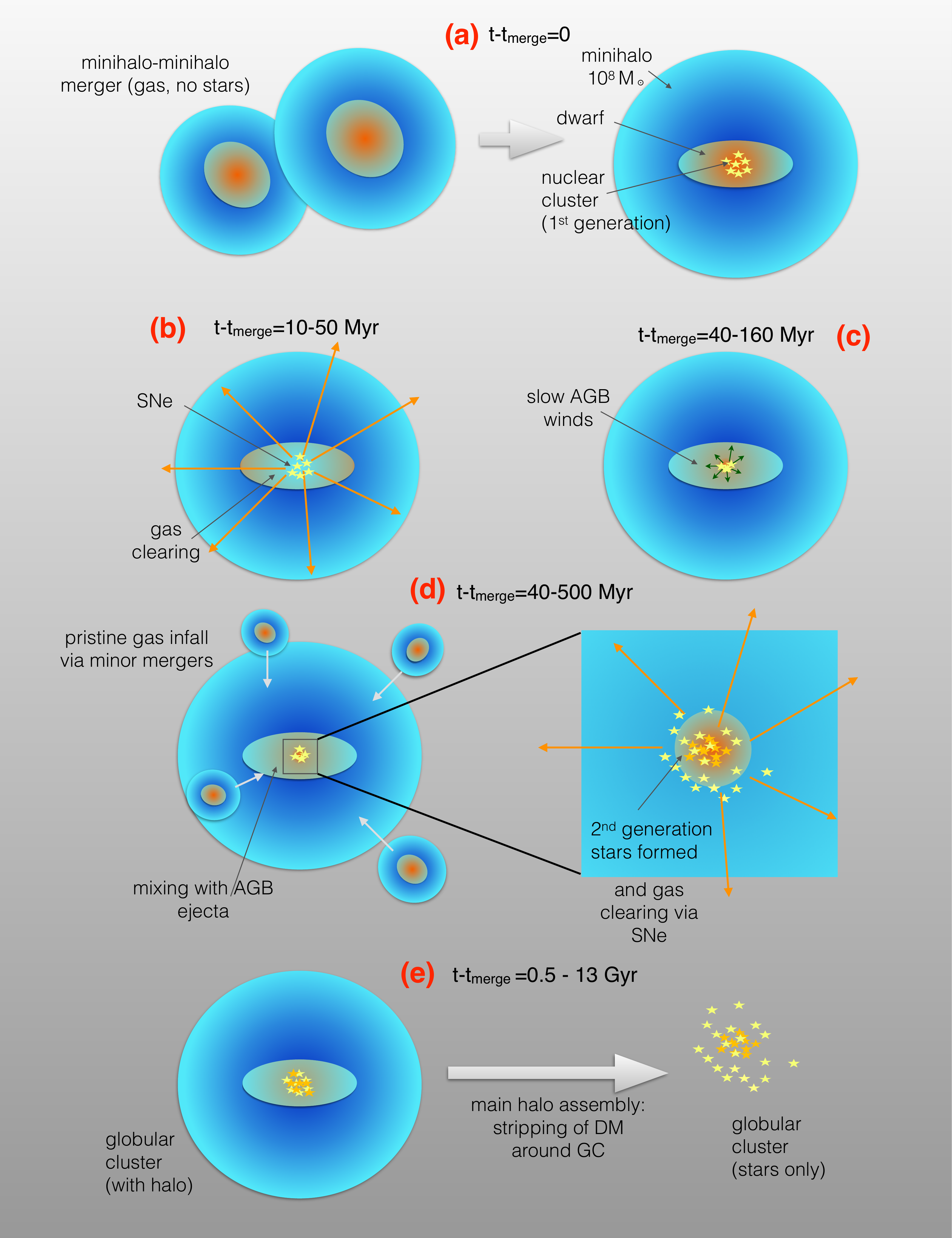}
\end{center}
\caption{{{Qualitative illustration of GC formation at high-$z$ through
  a minihalo-minihalo merger (Sec.~\ref{sec:model}).}}}\label{fig:cartoon}
\end{figure*}

We assume that compact star clusters form during major mergers of two
gas-rich and star-poor minihalos, whose gas has been previously
enriched by outflows from nearby halos to
$Z\gtrsim 10^{-2}~ Z_{\sun}$. The combined minihalo mass needs to
exceed the threshold for $HI$ cooling
($T_{vir}\gtrsim10^4~\mathrm{K}$), but if either of the progenitors
was above the cooling mass threshold, it must not have started forming
stars yet. Under these conditions, the merger will shock-heat and
concentrate the gas, leading to efficient formation of a star
cluster. Following the merger-triggered starburst, Type II supernovae
clear the gas remaining in the minihalo (escape velocity is a few tens
km/s), suppressing further star formation after
$\sim10~\mathrm{Myr}$. Subsequently, slow winds from AGB stars produce
chemically enriched gas \citep{dercole2008}, which is retained in the
potential well, and diluted by infall of new pristine gas through
minor mergers and/or accretion. This gas loading triggers one or more
bursts of star formation, more centrally concentrated than the first
generation, with merger mass ratios and timing after AGB pollution
imprinting diverse chemical signatures in individual clusters,
consistent with observations \citep{renzini2008}. The GC minihalo is
eventually incorporated into the merger-tree main branch, with
continued mergers and tidal interactions stripping DM and leaving a
``naked'' GC by $z=0$. {{Qualitatively, the scenario is illustrated in
Figure~\ref{fig:cartoon}.}}

\section{Quantitative modeling}\label{sec:simulation}To evaluate the plausibility of the scenario depicted in
Section~\ref{sec:model}, and compute the GC formation rate, their
ages, and spatial distribution at $z=0$ we resort to a DM-only
cosmological simulation run with Gadget2 \citep{springel2005}, with
setup described in \citet{trenti2009b,trenti2010_enrich}, and
\citet{ramirez-ruiz2014}, but tailored volume/mass resolution and
updated cosmology ($\Omega_{\Lambda,0}=0.685$, $\Omega_{m,0}=0.315$,
$\Omega_{b,0}=0.0462$, $\sigma_8=0.828$, $n_s=0.9585$, $h=0.673$;
\citealt{planck}). We simulate a $10^3~\mathrm{Mpc^3}$ (comoving)
volume from $z=150$ to $z=0$ using $N=403^3$ particles, while we run a
high-resolution version of the same initial conditions to $z=5.5$
($t_{lookback}\sim12.8$ Gyr) with $N=812^3$ particles (mass resolution
$8.3\times10^4~\mathrm{M_{\sun}}$\footnote{This guarantees that GC
  minihalos are well resolved with $N\gtrsim500$ particles, sufficient
  to characterise their properties \citep{trenti2010convergence}.}).
Our multi-scale approach is based on assigning membership of
high-resolution particles to $z=0$ halos through mapping of particles
IDs, and studying halo assembly at higher redshift from the full
resolution run, saving snapshots at uniform intervals of
$10~\mathrm{Myr}$.

\begin{figure}
\begin{center} 
\includegraphics[scale=0.43]{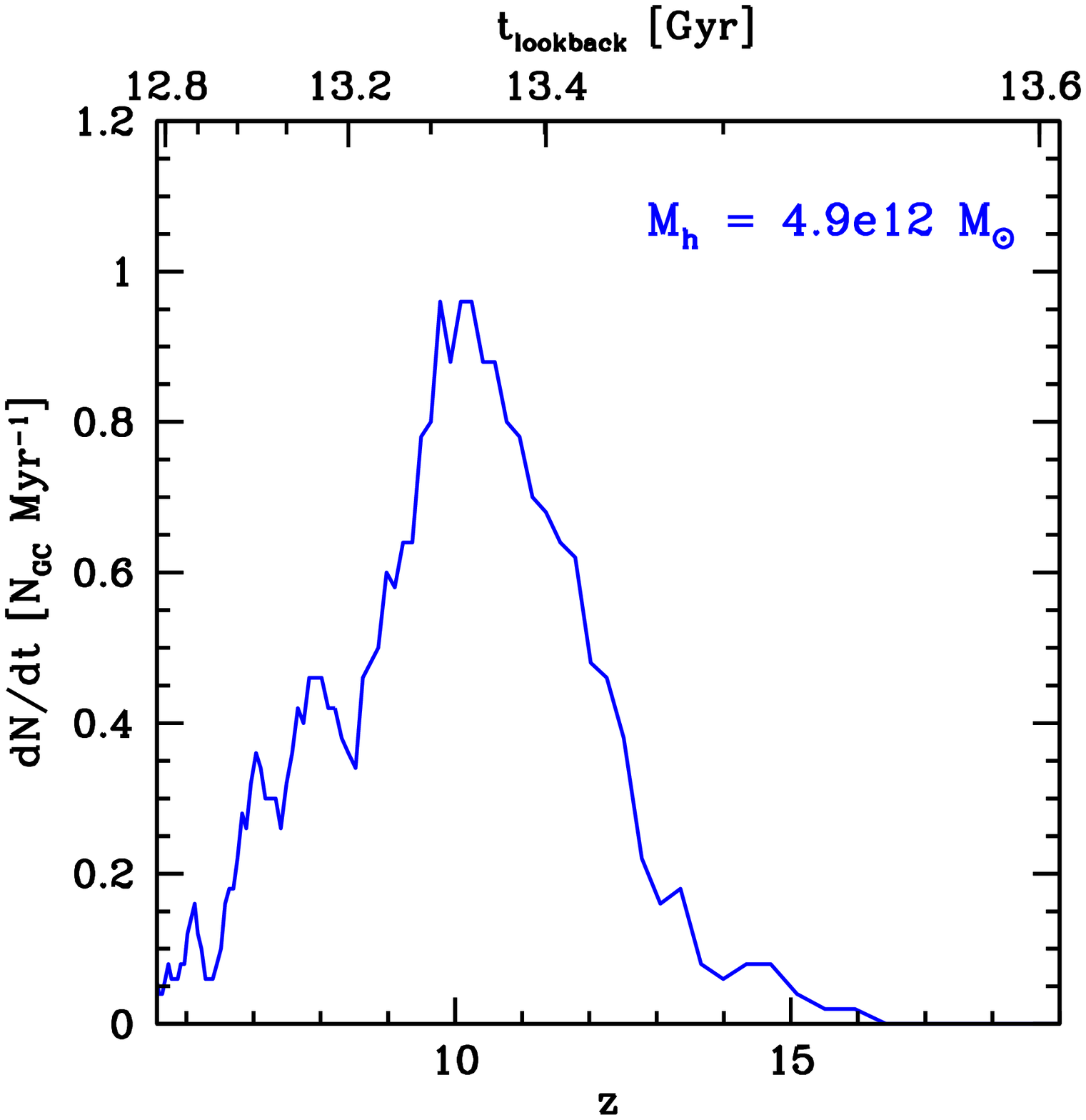}
\includegraphics[scale=0.43]{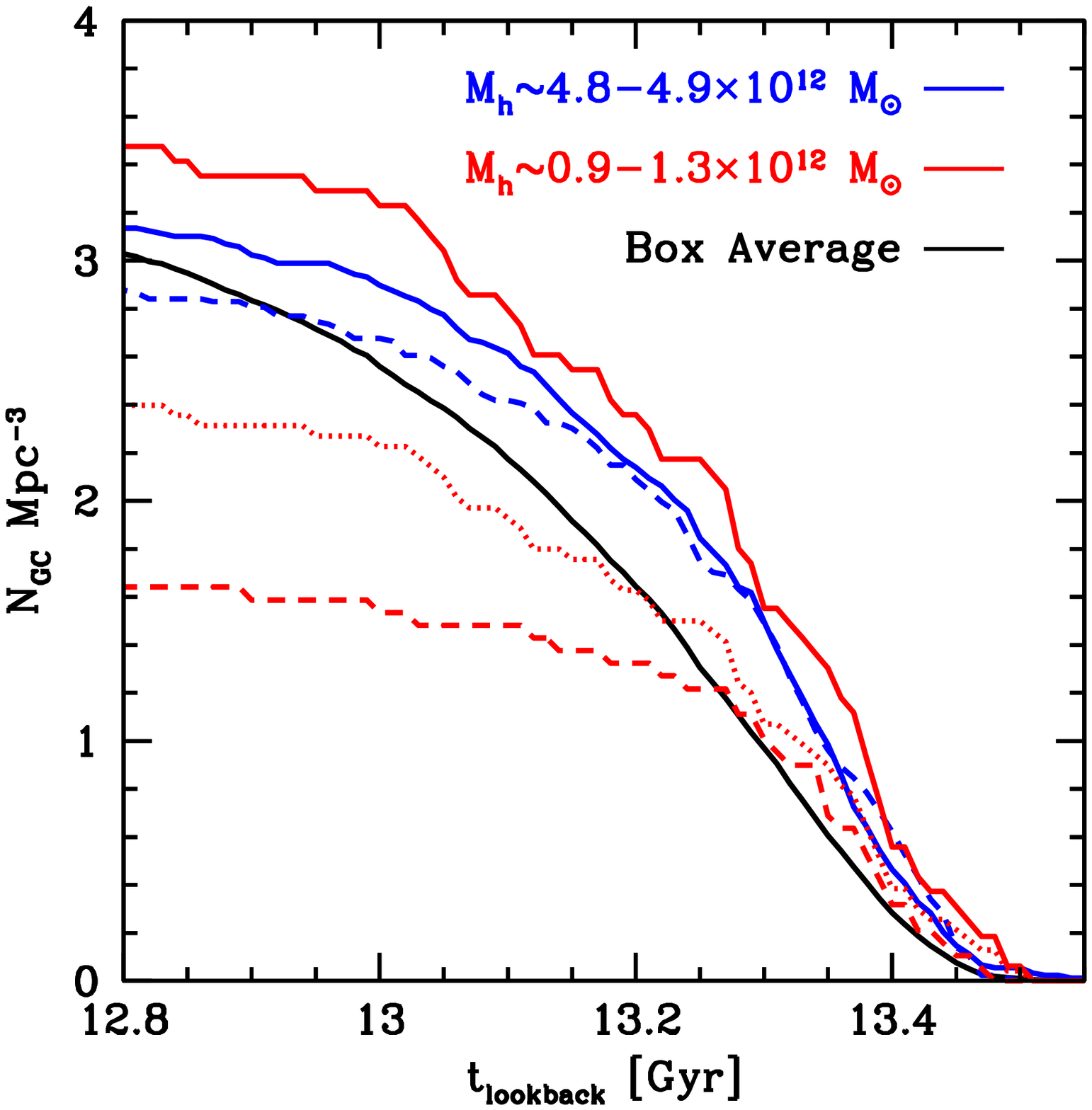}
\end{center}
\caption{Top panel: Formation rate 
  of GCs through minihalo-minihalo mergers in the most massive dark
  matter halo of our simulation
  ($M_h=4.9\times 10^{12}~\mathrm{M_{\odot}}$ at $z=0$). Data have
  been smoothed (boxcar filter width $50$ Myr). At
  $z>5.5$, $N=279$ clusters are formed with
  $\langle z\rangle=9.3$ (dispersion $\Delta z=1.9$). Bottom panel:
  Cumulative number of GCs formed per unit volume ($Mpc^{-3}$
  comoving) versus lookback time, for the two most massive halos of
  the simulated volume (solid/dashed blue), for the next three
  halos in mass ranking (solid, dotted and dashed red), and
  box-averaged value (black).}\label{fig:rate}
\end{figure}

The high resolution DM-only simulation is post-processed with the star
formation and chemical enrichment model of
\citet{trenti2009b} to flag minihalos that are in
gas-rich, star-free conditions. Specifically, we follow the formation
of Population-III (metal-free) stars in halos above the minimum
threshold for $H_2$ cooling with a time-evolving Lyman-Werner
background, triggering Population-III star formation when a pristine
halo crosses the mass $M_{min}(z)$ shown in Figure~1 of
\citet{trenti2009b}. Under these conditions, a halo is flagged as
enriched by Population-III stars, and then starts forming Population
II (metal-enriched) stars once above the hydrogen cooling limit
($T_{vir}>10^4~\mathrm{K}$). As basic description of metal pollution
within our DM-only framework, Population-II halos, once they begin
forming stars, are assumed to have outflows propagating with spherical
symmetry at fixed speed $v_{wind} = 60~\mathrm{km/s}$ (appropriate for
dwarf-like galaxies). Halos that grow with redshift staying below
$M_{min}(z)$ until $z\lesssim 13$ (point at which $M_{min}(z)$
corresponds to atomic cooling) may either form a ``late-time''
Population-III star, or a Population II cluster/dwarf galaxy if
polluted by outflows from nearby (proto)galaxies
\citep{trenti2010_enrich}.

As zeroth order characterisation of metal enrichment, we track for
each minihalo the number $\xi$ of nearby halos that have metal
outflows active for long enough to reach its center. We consider
outflow-enriched (pollution counter $\xi \geq 1$), star-free halos
with $T_{vir}\sim 10^4\mathrm{K}$ as potential GC birth-sites. The
condition that needs to be satisfied in our framework to create a GC
is a major merger with an another smaller
outflow-enriched\footnote{Mergers between a metal-enriched and a
  metal-free, star-free minihalo would also satisfy these requirements
  but such configurations are rare since pairs of
  merging minihalos have strongly correlated $\xi$.}  minihalo that
increases the mass so that $T_{vir}> 10^4~\mathrm{K}$. If either
progenitor has $T_{vir}\geq10^4\mathrm{K}$ pre-merger, then the merger
must happen before stars are formed. Otherwise supernova feedback
clears the minihalo of gas. Assuming a typical cooling time of
$(1-2)\times10^8$ yr (\citealt{tegmark1997}), we define a
timescale $\Delta t=150~\mathrm{Myr}$ from crossing of the cooling
threshold ($T_{vir}\geq10^4\mathrm{K}$). After such time we no longer
consider a GC as the outcome of a major minihalo merger. Furthermore,
we define the minimum mass ratio between progenitor halos as
$M_2/M_1>1/4$.

Finally, we save IDs for all particles involved in GC formation to
track halo membership, check for subsequent mergers, and characterize
their spatial distribution at $z=0$ through ID re-mapping into the
lower resolution run.

{{Our simplified gas treatment from DM-only simulations ignores the
    relative velocity difference between baryons and CDM imprinted at
    recombination time, which can significantly impact formation of
    high-$z$ halos with
    $10^3~\mathrm{K}\lesssim T_{vir}\lesssim10^4~\mathrm{K}$,
    suppressing the halo mass function
    \citep{tseliakhovich2010,naoz2011}, and the gas fraction
    \citep{naoz2013}. Quantifying the impact of stream velocity for GC
    formation is difficult without a hydrodynamic simulation, because
    of competing effects. Gas-rich minihalo mergers require
    progenitors where Population-III star formation is suppressed. In
    our framework this happens because of $H_2$ photodissociation, but
    stream velocity works as well. In addition, the halo bias would be
    higher \citep{tseliakhovich2010}, enhancing the merger
    probability. Overall, the number of gas-rich minihalo mergers may
    be only moderately suppressed, or possibly even enhanced, in
    contrast to the $\sim50\%$ suppression of star formation inferred
    for the general population of minihalos \citep{bovy2013}. We
    expect our estimate on the GC birth-rate to be accurate within a
    factor two or better compared to a full treatment of stream
    velocity.}}

\begin{figure*}
\begin{center} 
\includegraphics[scale=3.2]{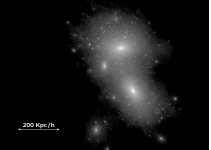}
\includegraphics[scale=3.2]{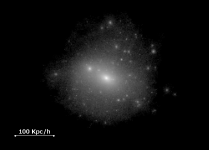}
\includegraphics[scale=3.2]{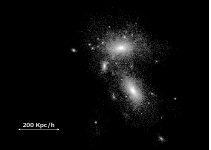}
\includegraphics[scale=3.2]{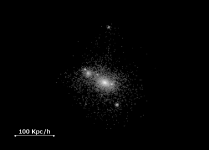}
\includegraphics[scale=3.2]{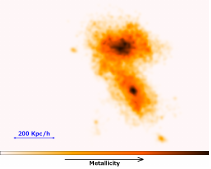}
\includegraphics[scale=3.2]{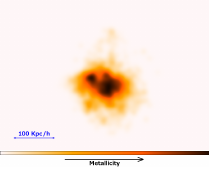}
\end{center}
\caption{Top-left panel: Projected spatial distribution of DM
  particles {{in the most massive simulated halo at $z=0$
  ($M_h=4.9\times10^{12}~\mathrm{M_{\odot}}$)}}. Bottom-middle panel:
  Spatial distribution (same redshift and same halo of
  top-left panel) of the subset of particles that have been part of
  the $N=279$ minihalo-minihalo mergers at $z>5.5$, which we propose
  are associated to the oldest GCs. Bottom-left panel: Projection of
  average value of chemical enrichment tag for the {{total
      metallicity of GCs}} (light to dark for increasing number of
  pollution events), highlighting increasing metallicity at small
  galactocentric radii {{in qualitative agreement with observations
      by \citet{marin-franch2009_ages}}}. Right column contains the
  same panels but for the fifth most massive simulated halo 
  ($M_{h}\sim8\times10^{11}~\mathrm{M_{\odot}}$).}\label{fig:rendering}
\end{figure*}

\subsection{Merger rate and age probability
  distribution}\label{sec:merger_rate}Figure~\ref{fig:rate} shows
predictions for the GC formation rate in the most massive simulated
halo ($M_h=4.9\times10^{12}~\mathrm{M_{\odot}}$, selected for best
statistics). At $z>5.5$ (Universe age $<1~\mathrm{Gyr}$), there are
$279$ minihalo-minihalo mergers fulfilling the conditions for GC
formation. The average merger redshift is $\langle z\rangle=9.3$
($t_{lookback}=13.2~\mathrm{Gyr}$) with dispersion $\Delta z=1.88$
(age spread $\sim200$ Myr). Minihalo-minihalo merging is strongly
peaked with redshift because of competing processes. At $z>12$ atomic
cooling halos are rare, and there is not sufficient time for chemical
pollution by outflows originating in nearby star forming sites. Later,
as the redshift decreases so does the merger rate
\citep{fakhouri2010}, and at $z\lesssim 8$ structure formation starts
shifting toward larger scales. This is qualitatively similar to the
circumstances leading to the ``Gamov peak'' in Nuclear Astrophysics.

Figure~\ref{fig:rate} is representative of the average GC formation
rate, albeit with significant halo-to-halo scatter. Its bottom panel
illustrates the cumulative GC number for the five most massive halos
in the box comprising a mass range
$(0.8-4.9)\times10^{12}~\mathrm{M_{\sun}}$, as well as the box
average. The number of GCs is normalised by the comoving volume of
each halo in the initial conditions (proportional to halo mass). The
box average normalization is the simulation
volume. Figure~\ref{fig:rate} depicts halo-to-halo variations of a
factor two. Furthermore, it shows that more massive halos (the five
blue and red lines) have slightly older GCs compared to less massive
halos (the box average line) by about $100$ Myr. This is a consequence
of the earlier chemical enrichment and biased halo mass function that
is found in overdense environments.

With an average time delay of $\sim 250$ Myr, the large majority of
the products of minihalo-minihalo mergers has a minor merger
(progenitor mass ratio $>1/10$) or receives a mass infall
$\geq 10^7~\mathrm{M_{\sun}}$ between two snapshots ($10$ Myr temporal
spacing). Major mergers are also frequent: $\sim 25\%$ of the first
generation clusters has one. More generally, from Extended
Press-Schechter modeling (see \citealt{trenti_santos08}) we derive
that a halo with $M_h=10^8~\mathrm{M_{\sun}}$ at $z=10$ will evolve
into a descendant that $500$ Myr later (by $z=5.8$) has doubled in
mass at $>98$\% confidence. Therefore, the high accretion rate on the
minihalos suggests that there is sufficient gas to mix, dilute
chemically the AGB ejecta from first generation stars, and concentrate
them at the halo center. These conditions can lead to formation of
multiple stellar populations \citep{carretta2009}, widely observed in
galactic GCs \citep{piotto2007}. However, detailed modeling requires
hydrodynamic simulations and is beyond the scope of this initial work,
focused on making predictions on the robust (parameter-free)
probability distribution of GC ages.

Our simulation does not include halos especially representative of the
MW/local group because of its small volume. Nevertheless, since
minihalo mergers are a universal process, we can rescale the results
to estimate the number of old galactic GCs. The MW halo mass is
debated, ranging from $\sim8\times 10^{11}~\mathrm{M_{\odot}}$ to
$\sim4\times10^{12}~\mathrm{M_{\odot}}$ at $95\%$ (e.g.,
\citealt{marel2012,phelps2013,kafle2014}), almost encompassing the
halos shown in Figure~\ref{fig:rate}. In these halos there are $279$
to $32$ old GCs, suggesting full consistency between observed
frequency and predictions in our framework (old GCs in the MW are
about $50\%$ of the total population, that is $N\sim80$). Overall, our
theoretical estimate of the old GC number is accurate within a factor
two, comparable to halo-to-halo variations. For improved estimates, GC
survival to $z=0$ is also important, as well as detailed modeling of
the hydrodynamics ({{stream velocity,}} impact of progenitor mass
ratio and cooling timescale $\Delta t$).

However, all these baryonic processes, {{including stream
    velocity}}, do not affect to first approximation the predicted age
distribution, which is the key observable that can falsify our
model. In this respect, we note that the GC ``formation epoch'' shown
Figure~\ref{fig:rate} is that of the minihalo-minihalo merger, and
thus an upper limit. Stars form in the following $\sim10^8$ yr (first
generation), and possibly a few hundred Myr later when further gas
accretion triggers subsequent generations. Assuming $2\times10^8$ yr
as typical delay for the average stellar age since the merger, our
best estimate of the absolute age of old GCs is $\sim13.0$ Gyr with a
$1\sigma$ spread of $\sim0.2$ Gyr. This is in full agreement with the
current determination of ages for the old GC population
$t_{age}=12.8\pm0.6$ Gyr \citep{marin-franch2009_ages}. We predict
that, if old GCs formed through mini-halo mergers, improvements in
absolute and relative age calibrations should converge toward GC
formation during the epoch of reionization and age scatter as small as
$\Delta t_{age}\sim0.2$ Gyr.

\subsection{GC masses,  galactocentric distribution and tidal
  stripping}\label{sec:checks}Our model quantitatively predicts the age distribution of
GCs formed through minihalo-mergers, and produces the correct order of
magnitude of objects (Section~\ref{sec:merger_rate}). To further
establish its plausibility, we investigate the expected GC
masses, galactocentric distribution and efficiency of tidal stripping
in removing DM around GCs by $z=0$.

A typical star formation efficiency used in studies of high-redshift
objects is $\epsilon_*=0.03$ (stellar to baryonic mass ratio;
\citealt{alvarez2012}). This gives for a minihalo with
$M_h\sim10^8~\mathrm{M_{\sun}}$ a stellar mass
$M_*\sim4\times10^5~\mathrm{M_{\sun}}$, consistent with the
average GC mass $\sim10^5~\mathrm{M_{\sun}}$ \citep{heggie_hut_book},
even if there is a diffuse stellar component formed during the merger
(and later stripped). Furthermore, our model has a cut-off scale given
by the HI cooling mass, naturally predicting a distribution of masses
peaked around a characteristic value (log-normal type), as observed
for GCs in both the Milky Way and external galaxies
(e.g. \citealt{harris1991,parmentier2005}). Thus, the model has no
need to invoke preferential disruption of lower mass star clusters
(e.g.  \citealt{fall2001}).

\begin{figure}\begin{center}\includegraphics[scale=0.4]{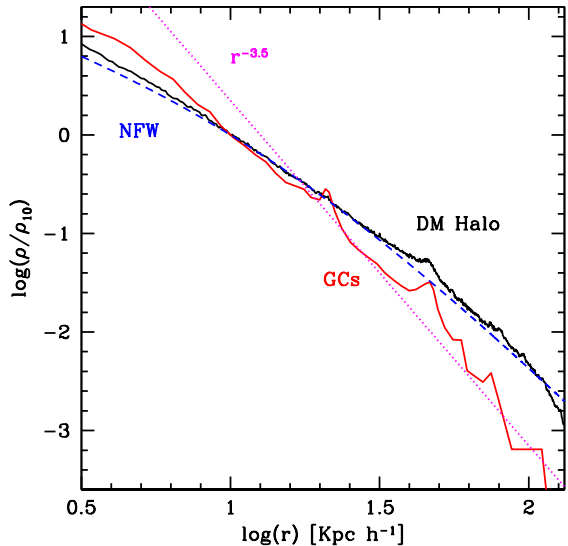}\end{center}
  \caption{Density profile versus galactocentric radius at $z=0$ for
    the halo shown in the right panels of
    Figure~\ref{fig:rendering}. The DM density (black-solid line) is
    well represented by a NFW profile (dashed-blue line), while
    particles associated to minihalo-minihalo mergers have a steeper
    profile (solid red), consistent with $\rho\sim r^{-3.5}$ observed
    in the distribution of Galactic GCs {{(dotted
        magenta)}}. Profiles have been normalized at
    $r=10~\mathrm{Kpc~h^{-1}}$.}\label{fig:density_profile}\end{figure}

Finally, we discuss model predictions for galactocentric GC
distribution and efficiency of tidal stripping of the DM envelope. We
tag in the $z=0$ snapshot the particles associated to GC formation in
the high-resolution, high-$z$ run. Results are shown in
Figure~\ref{fig:rendering}: tagged particles (middle panel) have
higher galactocentric concentration compared to all particles (top
panel). The {{$z=0$}} radial density profile of GC particles is
shown in Figure~\ref{fig:density_profile} for a {{typical
    non-interacting}} halo, demonstrating that it follows the slope
$\rho(r)\sim r^{-3.5}$ observed in the distribution of MW GCs. This is
not surprising since minihalo mergers are $\gtrsim2~\sigma$ DM peaks
at $z\sim9$, which \citet{moore2006} found to be distributed with a
$\rho(r)\sim r^{-3.5}$ profile by $z=0$. The middle panels of
figure~\ref{fig:rendering} also show a diffuse distribution of
particles involved in minihalo mergers. This supports our proposal
that tidal interactions during the merger process strip GCs of their
initial DM halos, although the results need to be confirmed by a full
high-resolution simulation to $z=0$. Last, through particle tagging we
derive the correlation between $z=0$ galactocentric distance and
chemical enrichment (Fig.~\ref{fig:rendering}, bottom panels). As
observed by \citet{marin-franch2009_ages}, we qualitatively obtain a
negative correlation between galactocentric radius and metallicity,
with GCs located centrally enriched by up to $\xi=15$ polluters, and
those in the outskirts having $\xi=1-2$.

\section{Conclusions}\label{sec:con}The observed age-metallicity distribution of galactic GCs shows a
bimodal population, with about half the objects residing along
an ``old'' branch with age $\sim12.8\pm0.6$ Gyr and spread comparable
to the relative age error
\citep{marin-franch2009_ages,forbes2010_age_metal}.  Here, we explored
a novel idea to form these old GCs at the center of DM minihalos with
virial temperature $T_{vir}\gtrsim10^4$ K
($M_h\sim10^8~\mathrm{M_{\sun}}$) and with the following additional
constraints:
\begin{itemize}\item Enrichment by metal outflows from nearby
  Population-II halos rather than by their own Population-III star
  formation;\item Major merger
  with a similar metal-enriched, but star-free minihalo within a time
  $\Delta t\sim150$ Myr from the crossing of the cooling threshold
  ($T_{vir}\gtrsim10^4$ K).\end{itemize}

The minihalo-minihalo merger is the key new ingredient to create a
compact star cluster. While the details of star formation depend on
complex baryonic physics that will be explored in future
investigations, we make robust and falsifiable predictions on the
probability distribution of the minihalo mergers (age of GCs), which
are discussed in Section~\ref{sec:merger_rate}. We obtain
$\langle t_{age}\rangle=13.0~\mathrm{Gyr}$ and
$\Delta t_{age}\sim0.2~\mathrm{Gyr}$, in agreement with current
observations.

Improvements in GC age measurements can uncontroversially falsify
the minihalo merger scenario. Also, model predictions may be within
reach of direct observations with the \emph{James Webb Space
  Telescope} (JWST). In fact, the number density of compact star
clusters at $z>7$ is sufficiently high ($1.5\times10^4$ arcmin$^{-2}$)
that JWST can identify some of them if highly magnified ($\mu\sim100$)
to $M_{AB}\sim-17$ by a foreground galaxy cluster. However, we note
that it will be challenging to discriminate against other compact
high-$z$ objects such as Population-III clusters (e.g.,
\citealt{zackrisson2012}).

Besides GC ages, other aspects of the framework {{such as multiple
    stellar populations formation triggered by minor mergers
    (Section~\ref{sec:model})}} are more speculative, since they
depend on a complex interplay between baryonic and DM physics. We
presented consistency checks in Section~\ref{sec:checks} to
characterize qualitatively the distribution of GCs, their metallicity
and stripping of the DM halo envelope, but further work is
required. In addition to detailed investigation of DM stripping,
perhaps one interesting aspect to follow-up is the connection
between GCs and dwarf galaxies. While earlier studies considered them
as two distinct classes in the luminosity-size plane, recent
work unveiled the existence of ultracompact dwarf galaxies with
intermediate properties (e.g., \citealt{jennings2014}). Indeed, our
framework predicts a continuum between classical dwarf galaxies (no
major mergers at formation, DM dominated) and old GCs (major merger,
DM envelope stripped by $z=0$). Overall, minihalo mergers appear to
provide a promising scenario to explore for the
formation of the oldest GCs observed in today's galaxies.

\acknowledgements It is our pleasure to thank Duncan Forbes, Diederik
Kruijssen and Sasha Muratov for useful discussions on an earlier
version of the manuscript, and an anonymous referee for helpful
comments.




\end{document}